\documentstyle[aps,epsfig,amstex,twocolumn]{revtex}

\begin{document}
\title{Effects of the trapping potential on a superfluid atomic Fermi Gas}

\author{G.\ M.\ Bruun}
\address{Nordita, Blegdamsvej 17, 2100 Copenhagen, Denmark}
\maketitle

\begin{abstract} 
We examine a dilute two-component atomic Fermi gas trapped in a harmonic potential in the superfluid phase.
For experimentally realistic parameters, 
the trapping potential is shown to have crucial influence on various properties of the gas. Using an 
effective hamiltonian, analytical results for the critical temperature, the temperature 
dependence of the superfluid gap, and the energy of the lowest collective modes are derived. These 
results are shown to agree well with numerical calculations. We furthermore discuss in more 
detail a  previous proposed method to experimentally observe the superfluid transition by looking 
at the collective mode spectrum. Our results are aimed at the present experimental effort to 
observe a superfluid phase transition in a trapped atomic Fermi gas. 
\end{abstract}

Pacs Numbers: 03.75.Fi, 05.30.Fk,

\
\
\

With the increasingly impressive experimental ability to  trap and cool atoms with fermionic 
statistics, a new and very active area within the field of ultracold atomic 
gases has emerged. Several groups are now involved in trapping and cooling atoms
with Fermi statistics and  temperatures
as low as $\sim 0.2T_F$ have been obtained for $^{40}$K and
$^6$Li~\cite{Experiments}. One major motivation for this experimental effort is to 
observe a phase transition to a superfluid state predicted to occur below a certain 
critical temperature $T_c$ for a two-component gas~\cite{StoofBCS}. 

In this paper, we examine various properties of such a two-component Fermi gas 
with an equal number of atoms $N_\sigma$ of mass $m$ 
in two hyperfine states $|\sigma=\uparrow,\downarrow\rangle$
trapped in a spherically symmetric harmonic potential of the form $U_0(r)=m\omega_T^2r^2/2$. 
Assuming that the gas is in the dilute regime, the interaction between atoms in the two 
different hyperfine states can be well described by a contact potential $g\delta({\mathbf{r}})$
with $g=4\pi\hbar^2a/m$ and $a$ the $s$-wave scattering length. The dilute regime is defined as $k_F|a|\ll1$
with $k_F=\sqrt{2m\mu_F}/\hbar$ and  $\mu_F$ the chemical potential.
For $a<0$ the effective atom-atom interaction is attractive and the gas is unstable toward the formation 
of Cooper pairs below a critical temperature $T_c$~\cite{StoofBCS}. However, the specific 
formula for the critical temperature was derived for a homogeneous system and is not necessarily valid 
for a trapped gas as we shall see. Indeed, 
it was shown in ref.\ \cite{BruunHeisel,HeiselBen} that the nature of the superfluid state depends on the 
coupling strength and the number of particles trapped. For many particles trapped and/or stronger coupling 
strengths (keeping $k_F|a|\ll 1$), the size of the Cooper pairs given by the coherence 
length $\xi_{BCS}=\hbar^2k_F/(\pi m\Delta)$, with $\Delta$ being the $BCS$ 
pairing gap (for $T=0$), is much smaller  than the size 
$R=\sqrt{2\mu_F/m}/\omega_T$ of the system. In this limit
the system can be described as a quasi-homogeneous bulk superconductor with a locally varying chemical
potential $\mu_F(r)=\mu_F-U_0(r)$~\cite{StoofBCS}. For weak atom-atom interactions or for few 
particles trapped on the other hand, 
the size  $\xi_{BCS}$ of the Cooper pairs is comparable to the size of the system. The 
 pairs are then formed between particles with time-reversed angular momentum quantum numbers 
$(l,m)$ and $(l,-m)$  residing in the same harmonic oscillator shell and  the gas is in the so-called 
intrashell regime~\cite{BruunHeisel}. 
It can in this limit in many ways be regarded as a large nucleus with some 
properties qualitatively different from the bulk regime. 
The transition between the two different regimes is roughly determined 
by $\xi_{BCS}\simeq R$ or equivalently $\Delta\simeq\hbar\omega_T$.
 Using the bulk theory prediction $\Delta=(2/e)^{7/3}\mu_F\exp(-\pi/2k_F|a|)$~\cite{gorkov} and 
the Thomas-Fermi result $\mu_F=(6N_\sigma)^{1/3}\hbar\omega_T$, the equation $\Delta=\hbar\omega_T$
yields $N_\sigma^*=\frac{1}{6}\left(\frac{e}{2}\right)^7\exp(3\pi/2k_F|a|)$. 
For $N_\sigma\ll N_\sigma^*$ atoms trapped, the gas is in the intrashell regime and for 
$N_\sigma\gg N_\sigma^*$ atoms trapped it is in the bulk regime.
With $k_F|a|=0.1$ this condition yields $N_\sigma^*=4\times10^{20}$ 
 and  for $k_F|a|=0.3$ we get $N_\sigma^*=9\times10^6$. Current experiments 
have $N\sim{\mathcal{O}}(10^5)$ particles trapped~\cite{Experiments} suggesting that in the dilute limit one is 
likely to be in the intrashell regime. The experiments 
seem to be aimed at  
using optical traps as the  confining potential since these offer more flexibility with respect to 
which atoms can be trapped~\cite{Experiments}. These optical traps  are 
rather tight with trapping frequencies $\omega_T\sim{\mathcal{O}}(10^3Hz)$ yielding  $\hbar\omega_T\sim{\mathcal{O}}(10^2nK)$.
Thus, the condition $k_BT_c\ll\hbar\omega_T$ [$k_BT_c\ll k_BT_F=(6N_\sigma)^{1/3}\hbar\omega_T$] for the intrashell regime 
is less restrictive than for magnetic 
traps. However, cooling schemes as the one described in \cite{Viverit},
 which enables the lowering of $T/T_F$ by orders of magnitude,  will probably have to be employed. 

For $k_F|a|\gtrsim0.3$ the gas is no longer in the dilute regime. Recently, a number of papers have been published examining 
the gas in this strongly correlated limit~\cite{Combescot}. It is presently unclear however, whether the gas is 
stable toward spinodal decomposition or other instabilities  and it is outside the scope of the present 
paper to describe the gas in this regime.

One property of great experimental importance 
is the critical temperature $T_c$ which we will now estimate for the gas in the intrashell regime. 
In this limit there is only pairing 
between particles with quantum numbers 
$(n,l,m)$ and $(n,l,-m)$ with the unperturbed  harmonic oscillator shell 
energy $\epsilon_n=(n+3/2)\hbar\omega_T$. This fact, combined 
with a number of semiclassical approximations  allows 
one to derive an effective Hamiltonian of the form~\cite{BruunHeisel}
\begin{gather}\label{EffectiveH}
\hat{H}_{eff}=\sum_{nlm\sigma}\xi_n \hat{a}^\dagger_{nlm\sigma}\hat{a}_{nlm\sigma}-\nonumber\\
\frac{2\hbar\omega_TG}{\Omega_{n_F}}\sum_{nlm,n'l'm'}(-1)^{m+m'}\hat{a}^\dagger_{nlm\uparrow}\hat{a}^\dagger_{nl-m\downarrow}
\hat{a}_{n'l'-m'\downarrow}\hat{a}_{n'l'm'\uparrow}
\end{gather}
where $\Omega_{n_F}=(n_F+1)(n_F+2)/2$ is the unperturbed degeneracy of the highest harmonic oscillator 
shell occupied with energy $\epsilon_{n_F}$ (ignoring the Hartree field for the moment),
 $\hat{a}_{nlm}$ is the operator removing a particle 
in the shell with energy $\epsilon_n$ and angular momentum quantum numbers $l,m$ and 
$\xi_n=\epsilon_n-\mu_F$. 
The effective coupling strength giving rise to the pairing correlations is given by $G=32\pi^{-2}k_F|a|/15$. 
From Eq.\ (\ref{EffectiveH}) a number of analytical results can be derived.  
The gap equation becomes~\cite{BruunHeisel}:
\begin{equation}\label{intragapeq}
   \frac{\Delta}{\hbar\omega_T}=G\sum_n^{n\le2n_F}\frac{\Delta}
  {E_n}[1-2f(E_n)] \,,
\end{equation}
where we have as a first approximation to the 
pseudopotential scheme introduced the cutoff $n\le 2n_F$. 
The quasiparticle energy is $E_n=(\xi_n^2+\Delta^2)^{1/2}$
and $f(x)=[\exp(\beta x)+1]^{-1}$ with $\beta=1/k_BT$. As
 $\Delta\ll\hbar\omega_T$ in the intrashell regime, shell effects are pronounced; the behaviour of the system 
depends crucially on whether the system is in a filled shell configuration or whether the highest occupied shell 
is only partially filled~\cite{BruunHeisel}. Since the filled shell configuration occurs only for certain ``magic'' 
numbers of particles trapped, we expect the typical experimental situation to be that of a partially filled 
highest shell [$\mu_F\simeq(n_F+3/2)\hbar\omega_T$].
 In this case, and for 
 $k_BT_c\ll\hbar\omega_T$ appropriate for the intrashell regime, Eq.\ (\ref{intragapeq}) yields the critical temperature 
\begin{equation}\label{Tceq}
\frac{k_BT_c}{\hbar\omega_T}\simeq\frac{G}{2-4\ln(e^\gamma n_F)G}=\frac{\Delta_0}{2\hbar\omega_T}.
\end{equation} 
Here $\gamma\simeq0.577$ is Euler's constant and $\Delta_0$ is the gap for $T=0$ in the intrashell regime~\cite{BruunHeisel}.
For $G\rightarrow0$, we have $k_BT_c/\hbar\omega_T=G/2$. 
In this very  weak interaction limit, there is only pairing in the partially 
filled shell right at the chemical potential ($n=n_F$) and the shells away from the chemical potential are unaffected 
by the pairing. For increasing $G$, the term $\ln(e^\gamma n_F)G$ in  Eq.\ (\ref{Tceq}) becomes important and 
$k_BT_c/\hbar\omega_T>G/2$. Physically, this corresponds to the interaction now being so strong that there is significant pairing 
in the shells away from the chemical potential; each Cooper pair is however still only formed between particles within the 
same shell. We see that $T_c$ in the intrashell regime given by Eq.\ (\ref{Tceq}) is very different from 
 $k_BT_c\propto\mu_F\exp(-\pi/2k_F|a|)$ relevant for the bulk regime~\cite{StoofBCS,gorkov}; for $G\ll 1$, 
Eq.\ (\ref{Tceq}) yields a critical temperature which is orders of magnitude higher than the bulk result. This 
stems from the fact  that the shell structure of the trap levels increases the density of states for certain energies. For 
$\ln[\exp(\gamma)n_F]G=1/2$, Eq.\ (\ref{Tceq}) diverges but this is clearly unphysical as the intrashell limit is characterized 
by $k_BT_c\ll \hbar\omega_T$. For such large values of $G$ the intrashell pairing ansatz is invalid and Cooper pairs are formed between 
particles residing in different shells.
It should be noted that Eq.\ (\ref{Tceq}) does not include the effect of induced interactions coming 
from density oscillations of the gas. In the homogeneous case, this effect is known to reduce $T_c$ 
by a factor $\sim2.2$~\cite{gorkov} and it should be investigated which influence this effect has in the intrashell
regime as well. However, it must be expected that the critical temperature including the induced interactions will be of the 
same order of magnitude as  Eq.\ (\ref{Tceq}) and thus very different from the bulk theory prediction. 

\begin{figure}
\centering
\epsfig{file=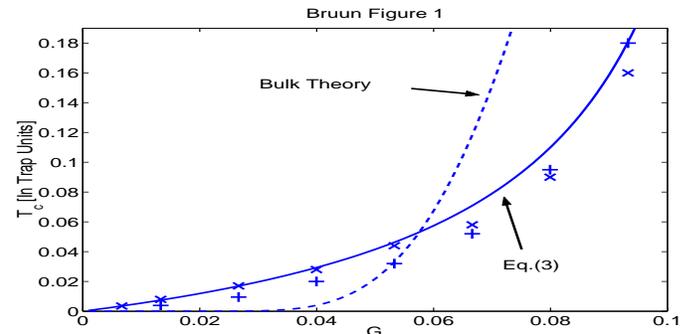,height=0.25\textwidth,width=0.5\textwidth}
\caption{The critical temperature $k_BT_c/\hbar\omega_T$ as a function of $G$. $+$'s and $\times$'s are with and without the
Hartree field respectively.}
\label{Tcfig}
\end{figure}

To check the validity of Eq.\ (\ref{Tceq}), we plot in Fig.\ \ref{Tcfig} $k_BT_c/\hbar\omega_T$  
as a function of $G$. The solid curve 
is Eq.\ (\ref{Tceq}) and the $\times$'s are obtained from a numerical calculation solving the Bogoliubov-de Gennes (BdG)
equations  derived from the full Hamiltonian using a pseudopotential scheme~\cite{BruunBCS}.
The dashed line depicts the bulk prediction $k_BT_c\propto\mu_F\exp(-\pi/2k_F|a|)$. 
We have chosen $n_F\sim30$ giving $N=2N_\sigma\sim10^4$ particles trapped. There is excellent agreement between the 
numerical results and Eq.\ (\ref{Tceq}) for $G\lesssim 0.05$. We conclude in agreement with ref.\ \cite{BruunHeisel}
that the intrashell description 
works well for $k_BT_c\ll\hbar\omega_T$.  In this 
limit on the other hand, the bulk theory is qualitatively wrong as expected. For stronger coupling, the nature 
of the superfluid state undergoes a transition and the intrashell description breaks down. There is now pairing between 
particles residing in different harmonic oscillator shells. This breakdown of the intrashell theory occurring for 
 $k_BT_c\gtrsim{\mathcal{O}}(\hbar\omega_T)$ for larger coupling strengths $G$ is not depicted in Fig.\ \ref{Tcfig}.
Eventually the coherence length $\xi_{BCS}$ becomes much smaller than the size of the system. As explained above, for 
$N\sim10^4$ trapped the gas is no longer in the dilute limit when $\xi_{BCS}\ll R$.

The Hartree field introduces a dispersion on the Harmonic oscillator levels with respect to the angular momentum $l$
lifting the shell degeneracy $\Omega_n=(n+1)(n+2)/2$. When $n_F\lesssim 30$ (depending on the size of the coupling strength), this 
splitting is small compared to the pairing energy $\Delta$ and it can be ignored~\cite{BruunHeisel}. For such systems, 
Eq.\ (\ref{Tceq}) for $k_BT_c$ should be accurate. For very large systems with  $n_F\gg 200$ 
on the other hand, the Hartree splitting of the shell is much larger than $\Delta$ and there is for weak coupling only 
pairing within states with a fixed $n$ and $l$~\cite{BruunHeisel}. In order to examine the effect of the Hartree field, 
 we plot in  Fig.\ \ref{Tcfig} as $+$'s $k_BT_c/\hbar\omega_T$  as a function of $G$ obtained from a numerical calculation
where the Hartree field is included. We see that effect of the Hartree field is to suppress $T_c$ for small $G$ 
due to the fact that the lifting of the shell degeneracy reduces the density of states at the chemical potential. For stronger 
coupling, the Hartree field increases $T_c$ as it in this limit enhances the spatial overlap between the orbitals of the 
Cooper pairs. However, from Fig.\ \ref{Tcfig} we see that 
Eq.\ (\ref{Tceq}) agrees qualitatively with the numerical results of the critical temperature. 
This is because $n_F\sim30$ is not quite in the limit where the Hartree field can be neglected, but also far away from
the regime $n_F\gtrsim200$ where there is only pairing within levels with a single $l$-value. 
Again, these results are in agreement with earlier calculations on the $T=0$ properties of the gas~\cite{BruunHeisel}.

\begin{figure}
\centering
\epsfig{file=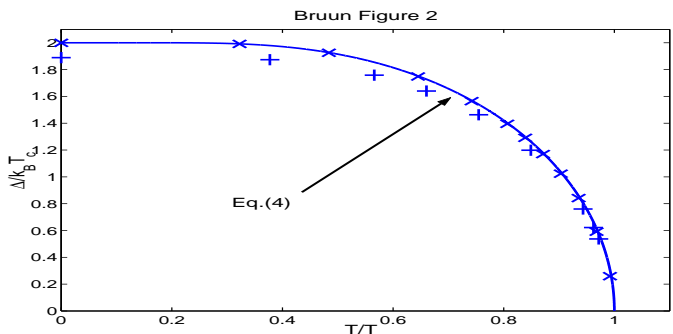,height=0.25\textwidth,width=0.5\textwidth}
\caption{The gap  $\Delta(T)/k_BT_c$ as a function of $T/T_c$. $+$'s and $\times$'s are with and without the
Hartree field respectively.}
\label{DeltaTfig}
\end{figure}

We now examine the temperature dependence of the pairing energy $\Delta$ in the intrashell regime. Solving 
Eq.\ (\ref{intragapeq}) for $\Delta(T)/\hbar\omega_T\ll1$, we obtain 
\begin{equation}\label{Tdependence}
   \frac{T}{T_c}=\frac{\Delta(T)}{k_BT_c\ln\left[\frac{1+\Delta(T)/2k_BT_c}{1-\Delta(T)/2k_BT_c}\right]}
\end{equation}
from which $\Delta(T)$ can be found by inversion. For $(T_c-T)/T_c\ll1$ such that $\Delta(T)/k_BT_c\ll1$ 
we have
\begin{equation}
\frac{\Delta(T)}{k_BT_c}=\sqrt{12}\sqrt{1-T/T_c}.
\end{equation}
In Fig.\ \ref{DeltaTfig}, we plot the lowest quasiparticle 
energy $E(T)/K_BT_c=\Delta(T)/k_BT_C$ (since $\xi_{n_Fl}=0$)  as a function of $T/T_c$. The solid 
line is Eq.\ (\ref{Tdependence}) and the $\times$'s and $+$'s are numerical results from a solution of the 
BdG Eq.\ with $\sim10^4$ particles 
trapped and $G=0.067$ where the effect of the Hartree field has been excluded for the $\times$'s.
We see that there is good agreement between the numerical results and  Eq.\ (\ref{Tdependence}) in the intrashell regime. 
In particular, we have $\Delta(T=0)/k_BT_c\simeq 2$  as was predicted in  Eq.\ (\ref{Tceq}). 
Again, the effect of the Hartree field does not change the qualitative behaviour of the system
for this set of parameters since $n_F\ll200$. So we conclude 
that Eq.\ (\ref{Tdependence}) describes $\Delta(T)$ well in the intrashell regime.

\begin{figure}
\centering
\epsfig{file=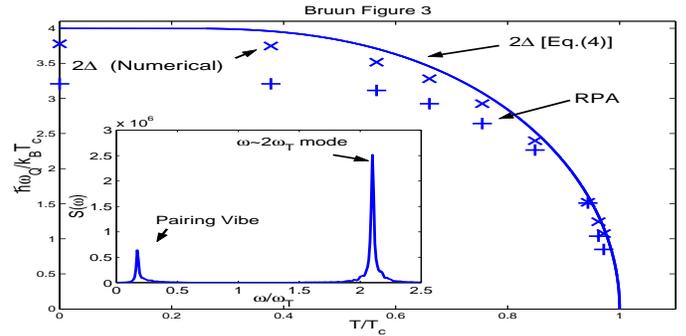,height=0.25\textwidth,width=0.5\textwidth}
\caption{The low energy quadrupole frequency $\hbar\omega_Q/k_BT_c$ as a function of $T/T_c$. 
The $+$'s are obtained from a numerical RPA calculation, 
the $\times$'s are $2\Delta/\hbar\omega_T$ from a numerical solution of the BdG eqn., and the line is $2\Delta/\hbar\omega_T$ 
with $\Delta$ given by Eq.\ (\ref{Tdependence}). The inset shows $S(\omega)$ (in trap units) as a function of $\omega/\omega_T$
for $T=0$.}
\label{Modefig}
\end{figure}

An important question  is how to experimentally observe the superfluid transition. One way 
is to look at the collective mode spectrum of the gas on which superfluidity should have 
several observable effects~\cite{BruunBencoll,Baranov}. 
When the gas is in the intrashell regime, two such effects
are  the emergence of low energy density oscillation modes of monopole and quadrupole symmetry.
The energy of these new modes should scale as $\hbar\omega\sim2\Delta\ll\hbar\omega_T$~\cite{BruunBencoll}. 
To examine this, we plot in Fig.\ \ref{Modefig} as $+$'s the frequency $\omega_Q$ of the low energy 
quadrupole mode as a function of $T$. The collective response is calculated using a RPA-scheme 
including the superfluid correlations as described in ref.\ ~\cite{BruunBencoll}. 
The solid line gives $2\Delta(T)$ obtained from 
Eq.\ (\ref{Tdependence}) and we also for comparison plot as $\times$'s $2\Delta(T)$ obtained from a numerical solution to the BdG eqn. The parameters 
used are the same as for Fig.\ \ref{DeltaTfig} and the effect of the Hartree field is included. 
We see that there is fairly good agreement between the numerical results from the RPA calculation and Eq.\ (\ref{Tdependence}). 
The energy of the collective mode given within the RPA is approximately $2\Delta$ indicating that the mode consists of 
breaking a Cooper pair. 
The fact that the  collective response frequency in general 
is lower than the pair breaking prediction $2\Delta$ can be understood as follows:
The excitation consists of the breaking of a Cooper pair of monopole symmetry costing $2\Delta$ in energy. However, the broken 
pair of particles then  form a new 
Cooper pair with  quadrupole symmetry; the difference $2\Delta-\hbar\omega_Q$ can 
be interpreted as the binding energy of the quadrupole Cooper pair. 
 The excitation can therefore be understood as the breaking of a Cooper pair 
with monopole symmetry and the subsequent formation of a Cooper 
pair of quadrupole symmetry, i.e.\ a pair vibration mode. 

The frequency $\omega_M$ of the lowest monopole mode is approximately the same as for the lowest quadrupole mode 
shown in Fig.\ \ref{Modefig}. Again, the excitation frequency is lower than the simple pair breaking prediction $2\Delta$ 
because the broken pair form a
new Cooper pair in an excited monopole state. We will in a future publication examine in more detail the binding energy of these 
excited Cooper pairs of monopole and quadrupole symmetry giving rise to the collective mode frequencies being somewhat lower 
than $2\Delta$. For the purpose of the present paper however, it is adequate to conclude that 
the onset of superfluidity below $T_c$ indeed does give rise to new density oscillation modes of monopole and quadrupole symmetry 
with  energy $\hbar\omega\sim2\Delta$ in the intrashell regime.

The collective modes can be excited by modulating the trapping frequency producing a perturbation with the 
right symmetry~\cite{Jin}.
There is however one important point concerning the monopole pair vibration mode: In order to excite 
this mode, one should apply a perturbation of the trapping potential 
 $\delta\hat{H}=\sin(\omega t)\sum_\sigma\int d^3rF({\mathbf{r}})\hat{\psi}_\sigma^{\dagger}({\mathbf{r}})\hat{\psi}_\sigma({\mathbf{r}})$ 
with $F({\mathbf{r}})\neq \alpha r^2$. This is because in the intrashell limit, the 
Cooper pairs are only formed between states within the same harmonic oscillator shell with unperturbed 
energy $(n+3/2)\hbar\omega_T$. Neglecting the
effect of the Hartree field, the states within one shell all have the same expectation value for $r^2$. Thus, 
due to the orthogonality 
between the eigenstates of $\hat{H}$, a perturbation  $\propto r^2$ (such as the modulation of the trapping frequency) 
will not probe  states which correspond to excitations of pairs 
within the same harmonic oscillator shell and the monopole pair vibration  mode $\hbar\omega_M\sim2\Delta$ will not be 
excited. To excite this mode, the scheme using off-resonant lasers demonstrated  might be useful~\cite{Ketterle}. 
For the numerical calculations, the monopole mode was excited by a perturbation with $F(r)\propto r^4$. 
A related issue is the spectral weight with which the pair vibration modes are excited as compared to the normal 
phase modes at $\omega\sim2\omega_T$. The spectral weight $S(\omega)\propto\sum_\nu|\langle0|\delta\hat{H}|0\rangle|^2
\delta(\hbar\omega-E_0-E_\eta)$ with $|0\rangle$ and $|\nu\rangle$ being the ground and excited states with energies
$E_0$ and $E_\nu$ respectively, is for $T=0$ in the quasiparticle approximation 
\begin{equation}\label{strength}
S(\omega)\propto\sum_{\eta \eta'}|\int d^3r [u_\eta v_{\eta'}+v_\eta
u_{\eta'}]F|^2\delta(\hbar\omega-E_\eta-E_{\eta'}),
\end{equation}
where the elementary quasiparticles with excitation energies $E_\eta$ are described
by the Bogoliubov wave functions $u_\eta({\mathbf{r}})$ and $v_\eta({\mathbf{r}})$~\cite{BruunJphys}. $\eta=(n,l,m)$
is shorthand for  the quantum numbers characterizing the quasiparticle. The collective pair vibration is predominantly built 
out of quasiparticle excitations with $n=n'=n_F$  whereas the normal phase excitation $\omega\sim2\omega_T$ 
is built out of excitations with $n=n_F-2,n_F-1,n_F$ and $n'=n_F+2$ or vice versa in Eq.\ (\ref{strength}). 
Now, for the $n=n_F$ shell, even a minute pairing will mix the particles and holes yielding 
$u_{n_Flm}({\mathbf{r}})=v_{n_Flm}({\mathbf{r}})=2^{-1/2}\phi^0_{n_Flm}({\mathbf{r}})$ where $\phi^0_{n_Flm}({\mathbf{r}})$
are the non-interacting harmonic trap eigenfunctions with energy $(n+3/2)\hbar\omega_T$~\cite{BruunHeisel}. 
This makes the $u_\eta v_{\eta'}$ in Eq.\ (\ref{strength}) large and depending 
on the spatial overlap between $F({\mathbf{r}})$ and the excitations involved, the strength of the pairing vibration 
will be of the same order as the normal phase excitation $\omega\sim2\omega_T$, although in general somewhat 
smaller since it is built from excitations from only the $n=n_F$ shell whereas 
the $\omega\sim2\omega_T$ mode is built from the shells $n_F-2,n_F-1,n_F$.
This is illustrated in the inset in Fig.\ \ref{Modefig}, where we plot $S(\omega)$ (calculated using the RPA) 
for $T=0$ with the parameters chosen above for the quadrupole 
mode excited by $F({\mathbf{r}})=r^2Y_{20}(\theta,\phi)$. We see that the
strength of the pairing vibration is comparable to the normal phase mode thereby making it experimentally observable. 

When the trap is not spherical, the single particle energies and wave functions are different making the quantitative 
predictions of this paper invalid. However, if all trapping frequencies $\omega_i$ ($i=x,y,z$) are such that 
$\hbar\omega_i\gg k_BT_c$, the qualitative conclusions of this paper still holds: $T_c$ can be orders of magnitude 
higher than the bulk result due to increased density of states for certain energies and there will be pair vibrations
with $\hbar\omega\sim2\Delta$. If, on the other hand, one frequency is loose (say $\hbar\omega_z\ll k_BT_c$), then the 
gas will be in some interesting mixture of the bulk and the intrashell regime which warrants further study. 

In conclusion, we have shown that for experimentally realistic parameters, the trapping potential has a significant influence on 
the properties of a trapped atomic Fermi gas in the superfluid phase in the dilute limit. In the intrashell regime, 
we used an effective Hamiltonian  to 
derive analytical results concerning the critical temperature, the temperature dependence of the pairing gap  and the lowest 
collective oscillations of monopole and quadrupole symmetry. These results were shown to agree well with numerical calculations. 
We also discussed in further detail a previously proposed scheme to detect the onset of superfluidity with decreasing temperature 
by looking at the collective mode spectrum. Much of the analysis of this paper should be relevant for the present experimental 
effort aimed at observing the superfluid transition in a two-component trapped atomic Fermi gas.

\

We acknowledge valuable discussions with C.\ Salomon, B.\ Mottelson, and J.\ R.\ Krumrine.

\end{document}